\documentclass[preprint,fleqn,showkeys]{revtex4}
\usepackage{xcolor,soul,framed} 

\colorlet{shadecolor}{yellow}
\usepackage[pdftex]{graphicx}

\usepackage[cmex10]{amsmath}
\usepackage{array}
\usepackage{mdwmath}
\usepackage{mdwtab}
\usepackage{eqparbox}
\usepackage{url}
\usepackage{tabularx}

\begin{document}
\title{Analysis of risk propagation using the world trade network}
\author{Sungyong \surname{Kim}}
\author{Jinhyuk \surname{Yun}}
\email{jinhyuk.yun@ssu.ac.kr}

\affiliation{School of AI Convergence, Soongsil University, Seoul 06978, South Korea}

\date[]{}

\begin{abstract}
An economic system is an exemplar of a complex system in which all agents interact simultaneously. Interactions between countries have generally been studied using the flow of resources across diverse trade networks, in which the degree of dependence between two countries is typically measured based on the trade volume. However, indirect influences may not be immediately apparent. Herein, we compared a direct trade network to a trade network constructed using the personalized PageRank (PPR) encompassing indirect influences. By analyzing the correlation of the gross domestic product (GDP) between countries, we discovered that the PPR trade network has greater explanatory power on the propagation of economic events than direct trade by analyzing the GDP correlation between countries. To further validate our observations, an agent-based model of the spreading economic crisis was implemented for the Russia-Ukraine war of 2022. The model also demonstrates that the PPR explains the actual impact more effectively than the direct trade network. Our research highlights the significance of indirect and long-range relationships, which have often been overlooked.
\end{abstract}


\maketitle

\section{Introduction}
An economy is an example of a complex system~\cite{schweitzer2009economic}. It can be represented by myriad agents ranging from individuals to larger units, such as states, industrial sites, or countries. However, owing to the complexity of an economic system, the majority of what we have observed is at the aggregate level, making it difficult to map the entire landscape. Connecting macroscale observations and microscale agents is a challenge, and we must therefore estimate the dynamic properties of such systems based on insights from complex systems, such as how an economic network can be used to comprehend the interaction and structural properties of economic agents.

Owing to the complex nature of the economy, crises that begin in one country can spread to the rest of the world. For instance, in most countries, the global economic system was halted by the COVID-19 pandemic in 2020~\cite{borio2020covid}. The failure of Lehman Brothers, one of the largest U.S. banks, triggered a global economic crisis in 2008~\cite{sieczka2011lehman}. The world is becoming increasingly entangled, and a crisis can be caused not only by large events but also minor events~\cite{jaaskela2010butterfly}. In other words, as globalization continues to advance, the impact of a small economic event in one country can spread throughout the world. 

To examine interactions at the nation level, we frequently employ the world trade network, which assumes that the degree of interdependence between two countries is proportional to their (relative) trade volume ~\cite{garlaschelli2005structure, de2011world}. A node in such networks represents a single country, and the link weights are proportional to the trade volume. The trade volume is commonly regarded as a measure of a nation’s dependence on other countries; however, economic events can be transmitted indirectly through a third country. Consider two countries, $A$ and $B$, with a weak coupling that interacts strongly with a third common country, $C$. The world trade network suggests that economic dependence between $A$ and $B$ should be negligible; however, there may be an unseen dependence between them by the shared strong trade partner $C$~\cite{yun2022generalization}. In the world trade network, the aforementioned relationships can be categorized as second-neighbor relationships, and higher-order interactions between shared partners also exist.

In this study, the world trade network serves as the underlying structure of the economic propagation between countries. We then estimate the total dependence, which includes both direct and indirect influences. We employ the personalized PageRank (PPR) algorithm~\cite{page1999pagerank}, which is a random walker-based method for determining the level of interdependence between countries and measures the probability that random walkers from one node will visit another (see Methods). Clustering and correlation analyses confirm that the two networks are comparable to a certain degree~\cite{traag2019louvain}. We also examined the importance of higher-order dependencies between remote nodes, as demonstrated by the correlation between the gross domestic product (GDP) coupling and the link weights. Our new agent-based model, considering the Russia-Ukraine war of 2022, indicates that the PPR has greater explanatory power than the world trade network, even when modelling is applied on a direct trade network~\cite{bonabeau2002agent}.

\section{Methods}

\subsection{Constructing the world trade network}\label{subsec:network_construction}
To investigate the economic interdependence of different countries, we constructed a world trade network. We retrieved international export data from the World Bank's World Integrated Trade Solution database ~\cite{monarch2017learning}. Such data include the annual volume of international trade. To build a single static network from yearly variances in trade volume, we took the average trade volume between 1990 and 2020. We also collected annual GDP data of 41 selected countries from the OECD Economic Outlook dataset between 1990 and 2020 (\url{https://www.oecd.org/economic-outlook/}). We only use countries considered by both the OECD and World Bank. In addition, to account for Ukraine and the Russian Federation, we used the KOSIS GDP dataset (\url{https://kosis.kr/}) for agent-based modeling, which are absent from the OECD dataset. We also use the average value between 1990 and 2020 for the KOSIS dataset.

Trade between countries can be represented as directed and weighted networks, with each node representing a node and the link weight being proportional to the trade volume. Owing to the significant differences in economic volume between countries, we apply summation-based normalization to gauge the dependency of a given country $i$ on other nations. We divide the trade volume for a link based on the total imports of the country in the following manner:

\begin{equation}
    \hat t_{ij} = \frac{t_{ij}}{\sum_k t_{ik}},
\end{equation}

\noindent where $t_ij$ represents the volume of raw imports from country $j$ to a given country $i$. This normalization rescales the edge strength of all countries to the same magnitude such that the total out-weight of a country is normalized to 1. After constructing the base world trade network, a random walker-based algorithm is used to build the total influence network. The details are presented in Section~\ref{subsec:ppr}.

\subsection{Personalized page rank as the total influence between countries}\label{subsec:ppr}
PageRank was proposed as an indicator of the importance of each webpage (node) based on the interconnection relationship between webpages ~\cite{page1999pagerank}. PPR, also known as a random walk with restarts, is a modified version of the original PageRank algorithm that measures the personalized significance of a target node $t$ regarding source node $s$. In a network analysis, the PPR similarity between two nodes represents the probability that a random walker starting from $s$ will visit $t$ and is a common measure of higher-order similarity~\cite{haveliwala2003topic}. In the world trade network, a random walker traverses from a specified source node $s$ with a probability proportional to the link weights outgoing to the neighbors of the current location of the walker. During this traversal, the walker returns to source node s with a probability of $1 \geq \alpha \geq 0$ (teleport probability). Likewise, the walker returns to source node $s$ if its current site has no links. The PPR can be calculated practically through a recursive multiplication of matrix $M$, which is defined as follows:

\begin{equation}
r = M \cdot r.
\end{equation}

\noindent Here, $M$ is a dense matrix in which an element $M_{ij}$ represents the PPR similarity from node $i$ to node $j$. Element $M_{ij}$ is defined as follows:

\begin{equation}
M_{ij} = 
\begin{cases}
(1-\alpha)M'_{ij} + \alpha, & \text{if } i = s \\
(1-\alpha)M'_{ij}, & \text{otehrwise},
\end{cases}\label{eq:ppr2}
\end{equation}

\noindent where $M_{ij}$ is assigned as $A_{ij}$, which is an element of the adjacency matrix, at the start of the recursive multiplications. Following the original study~\cite{page1999pagerank}, we employed the teleport probability $\alpha = 0.15$. Note that PPR similarities cannot be obtained if there is no path between two nodes. In this study, the PPR was computed using the Python-based package \texttt{fast-pagerank} retrieved from \url{https://pypi.org/project/fast-pagerank/}. We hereafter use the term PPR for indicating a network constructed using the PPR algorithm, unless otherwise specified.

\subsection{Modularity based clustering: Leiden Algorithm}\label{subsec:leiden}
We identify the cluster for a given world trade network using the Leiden algorithm~\cite{traag2019louvain}, a refined version of the Louvain algorithm~\cite{blondel2008fast} that supports numerous modularity and quality functions. We also employed the quality function $Q$ of the Potts model under the Reichardt and Bornholdt's configuration null model (RB model)~\cite{leicht2008community} as follows:

\begin{equation}\label{eq_quality_function}
Q = \sum_{ij} \left(A_{ij} - \gamma \frac{k_i k_j}{2m} \right)\delta(\sigma_i, \sigma_j),
\end{equation}

\noindent where $A_{ij}$ is an element in the adjacency matrix denoting the weight of the link between nodes $i$ and $j$. The node strengths $k_i$ and $k_j$ are defined as the total sum of the outgoing and incoming link weights of nodes $i$ and $j$, respectively. In addition, $m$ represents the total number of nodes in the network. An adjustment of the resolution parameter $\gamma$ alters the number of clusters. A larger $\gamma$ value yields a more fine-grained cluster. In this study, we used $\gamma = 1$, which is identical to the original Newman-Girvan modularity ~\cite{newman2004finding}. A parameter $\sigma_i$ specifies the community assigned to node $i$, with the Kronecker delta $\delta(\sigma_i, \sigma_j)$ being 1 if $\sigma_i = \sigma_j$, and 0 otherwise.

\section{Results}

\subsection{Clustering analysis of world trade networks}
Our primary objective is to determine whether it is necessary to consider higher-order influences when analyzing the economic dependence between countries. To determine whether a visible difference exists, we begin by conducting a clustering analysis on the constructed networks. 

\begin{figure}[h!] 
\centering 
\centerline{\includegraphics[width=\textwidth]{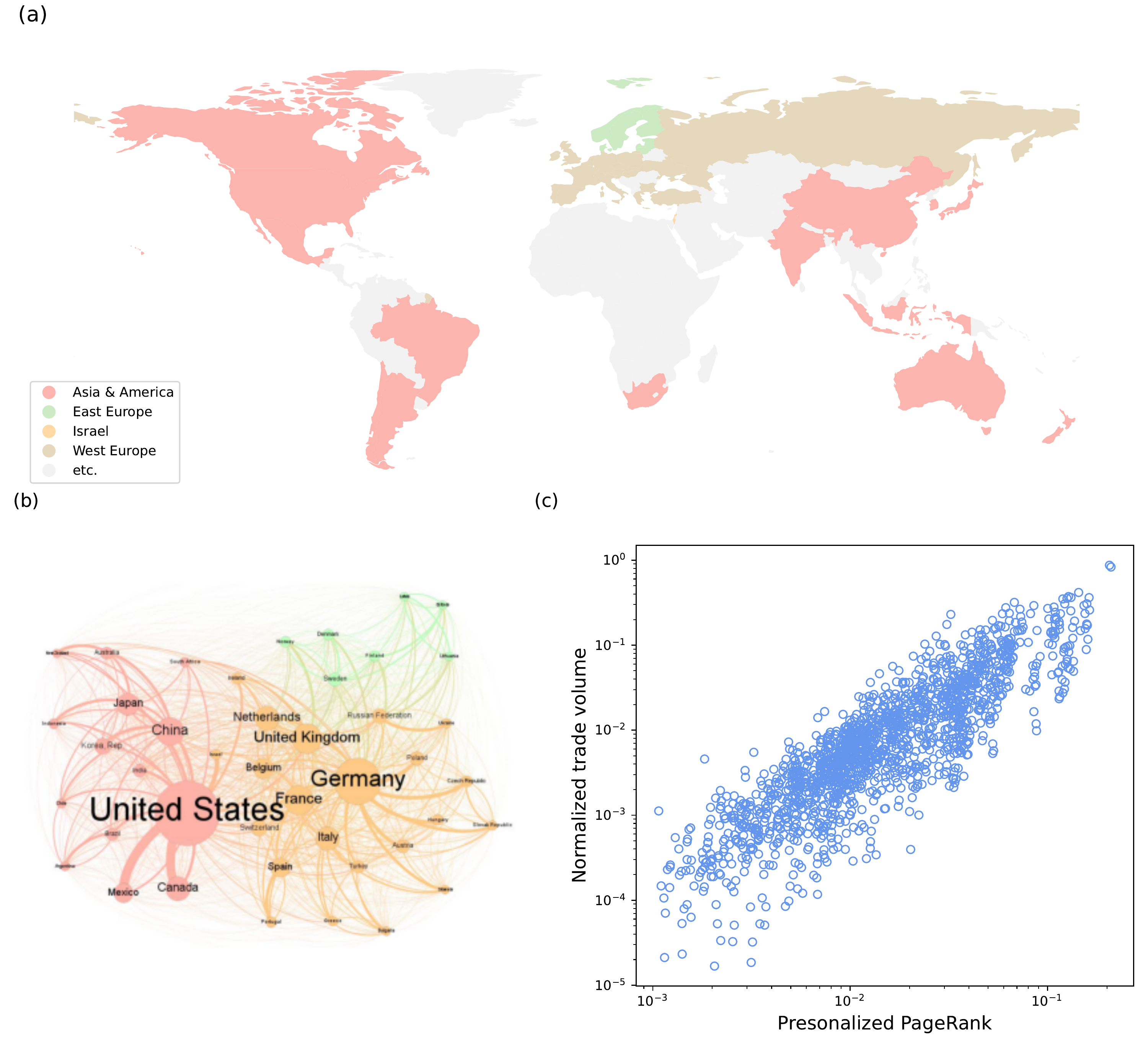}}
\caption{Clusters of world trade networks: (a) clusters displayed on a geographical map, and (b) clusters presented as a network representation. The clustering outcomes are produced using the Leiden algorithm (see Section~\ref{subsec:ppr}). (c) The correlation between the raw (normalized) trade volume and the total influence network as measured using the PPR.} 
\label{fig:ClusteringMap} 
\end{figure}

Because both the PPR and normalized trade volume are also based on a raw trade network, they should reflect the structure of a world economic exchange, and should therefore be related to a certain degree. Herein, we present evidence indicating that PPR is correlated with a normalized trade network. First, we focus on the fact that the PPR provides nearly identical clusters to such a network (see Fig.~\ref{fig:ClusteringMap}(a) and (b)). For both networks, we yielded three different clusters: Asia and America, Western Europe, and Eastern Europe. All countries except Israel are classified as equivalent clusters for both networks. Israel is classified as part of the Asia and America cluster for the normalized trade network and the Western Europe cluster for the PPR network. 

We then concentrate on the interdependence between two node pairs (countries) in which raw trade exists. We discovered a positive correlation between the PPR interdependency and the normalized trade network (see Fig.~\ref{fig:ClusteringMap}(c)). In particular, the correlation between the PPR and normalized trade network is correlated to each other regrading the Spearman rank correlation ($\simeq 0.78$).

\begin{figure}[h!] 
\centering 
\centerline{\includegraphics[width=\textwidth]{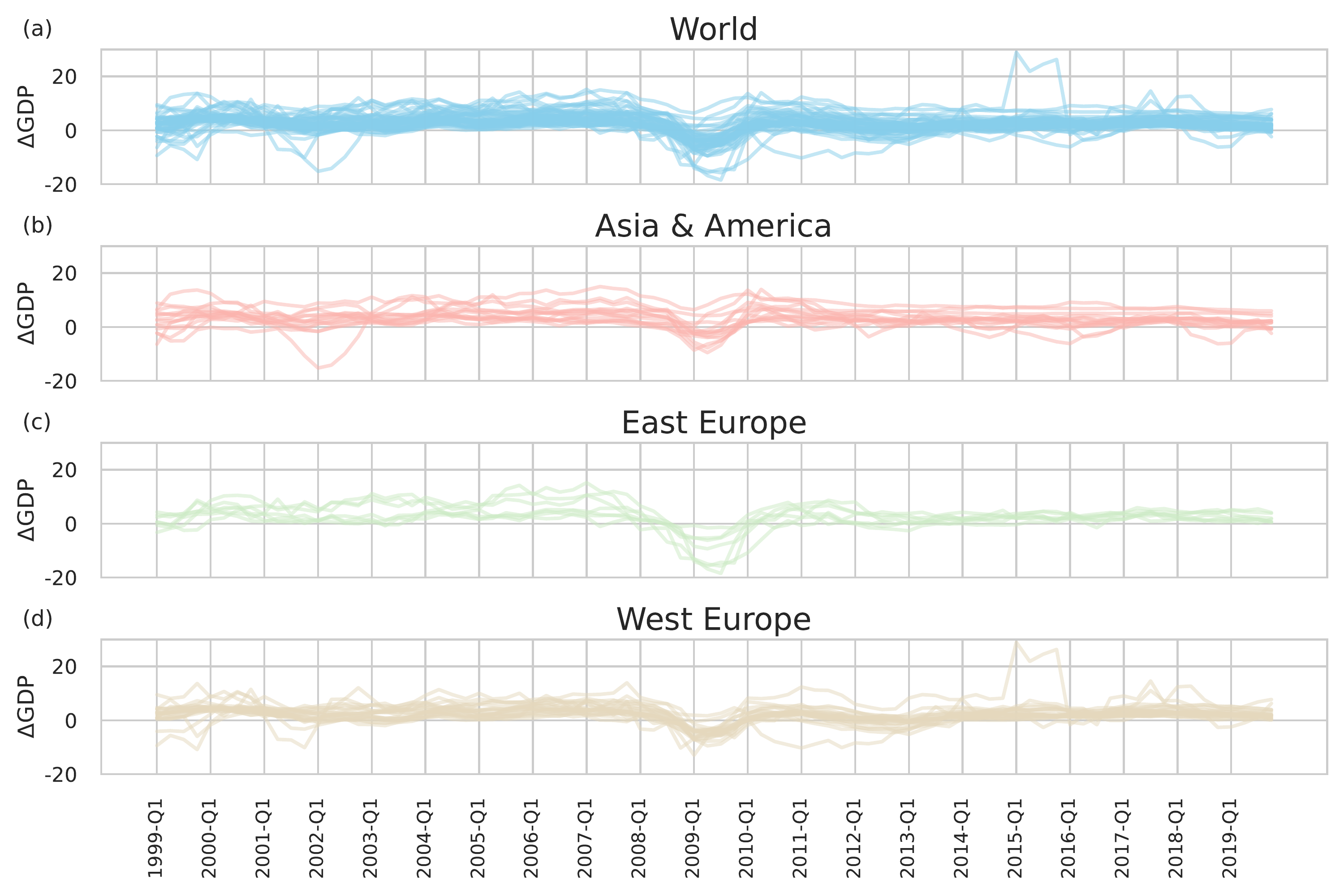}}
\caption{Quarterly difference in GDP for countries based on the clusters shown in Fig.~\ref{fig:ClusteringMap}: (a) Worldwide, (b) Asia and America, (c) Eastern Europe, and (d) Western Europe}
\label{fig:ClusteringGDP} 
\end{figure}

We also examined the correlation between GDP fluctuations to determine whether such trade networks influence real economic fluctuations (Fig.~\ref{fig:ClusteringGDP}). The global economy is thus interdependent. Consequently, all countries fluctuate when there is a significant increase or decrease in GDP. For example, every country experienced a decline in GDP during the 2009 global economic crisis. However, the degree of change in GDP varies by country, indicating that the impact of the economic crisis varies based on the economic structure of the country or its major trading partners. We also measured the similarity of the GDP time series between countries to numerically verify the entanglement using the dynamic time wrapping (DTW) technique~\cite{berndt1994using}, which is frequently applied to estimate the similarity between different time series. A higher DTW similarity indicates a greater association between time series. We observed a global positive correlation between countries (0.68 regarding DTW similarity). In addition, we found that the similarity between each cluster was greater than the global average (0.77 for cluster 1, 0.71 for cluster 2, and 0.69 for cluster 3). Although we used the cluster derived from the PPR, the results for the normalized trade volume are nearly identical because the clusters are similar, except for Israel. A stronger correlation suggests that the degree of economic interdependence is associated with the influence of the trade networks.

In summary, the global economy is interrelated and the degree of interdependence is related to the trade volume. In the following analysis, we demonstrate the difference in the explanatory power of the economic association between the PPR and normalized trade volume.

\subsection{Degree of economic association and two types of world trade networks}\label{subsec:correlations}

\begin{figure}[h!] 
\centering 
\centerline{\includegraphics[width=\textwidth]{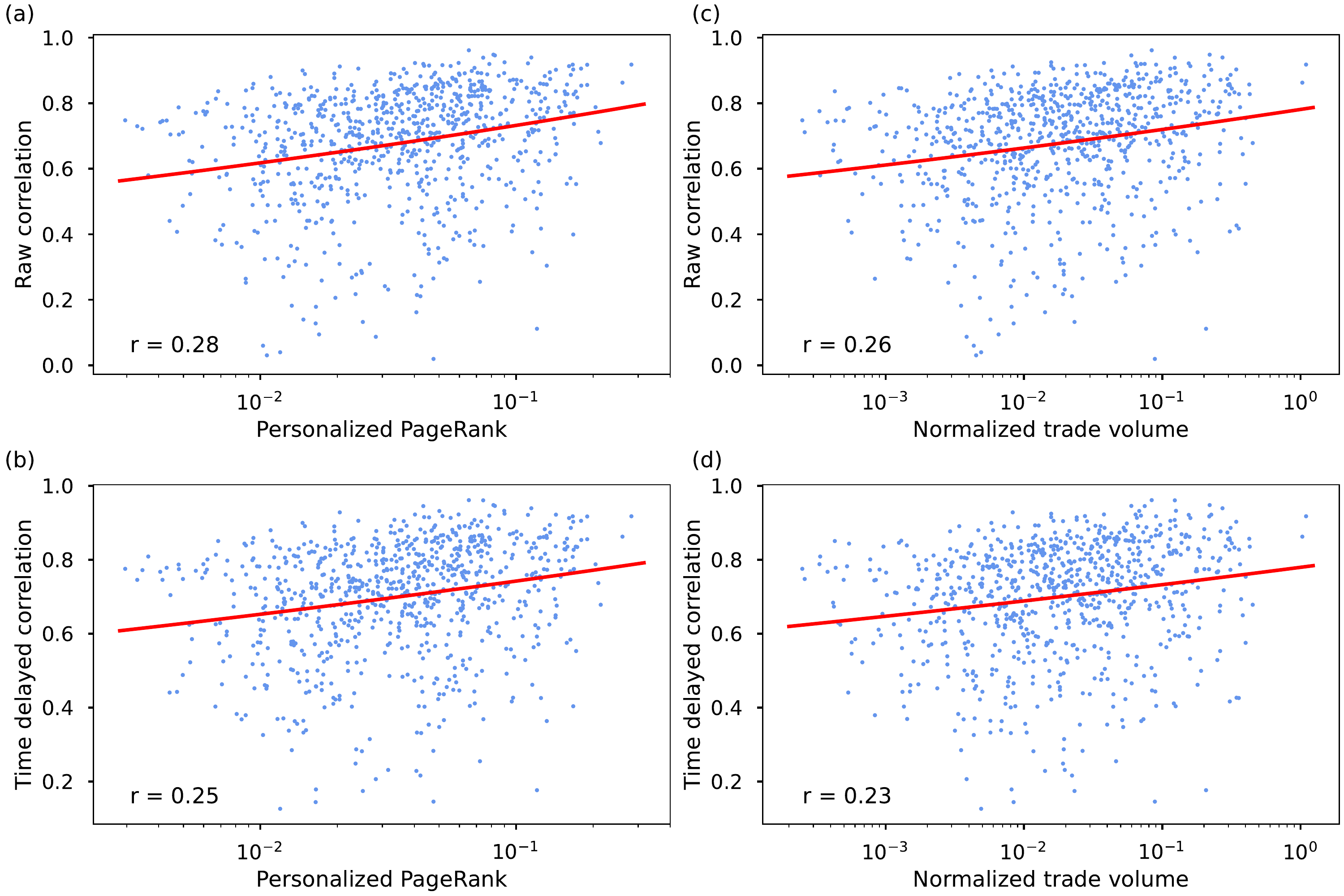}}
\caption{Degree of associations between weights of the world trade networks and the correlation of the GDP time series: (a) PPR with a raw correlation, (b) PPR with the maximum time-delayed correlation, (c) normalized trade volume with a raw correlation, and (d) normalized trade volume with the maximum time-delayed correlation. Here, correlation between time series are measured with cross-correlation using \texttt{scipy.signal.correlate}~\cite{plerou1999universal}. The $r$ value for each panel denotes the Spearman rank correlations of the scatter to present the degree of association. For all panels, the horizontal axis is the average of the bidirectional link weights between two nodes for each network.} 
\label{fig:Correlation} 
\end{figure}

To date, we have focused primarily on clusters derived from global trade networks. A significant advantage of the PPR is its ability to incorporate both direct and indirect influences between distant nodes. Although the world trade network is dense and interconnected, the trade volume itself is highly heterogeneous, resulting in a wide range in the degree of influence (Fig.~\ref{fig:ClusteringMap} (c)). The next logical step is to determine the role of indirect influence on global trade networks. In other words, it is necessary to investigate the long-range interactions between nodes with negligible trade–volume correlations.

The preceding results raise the question of whether the PPR captures the degree of economic association between two countries better than the normalized trade volume. This hypothesis is examined by comparing the correlation between the degree of GDP association and the weights of two distinct types of network. To determine the extent of associated GDP between countries, we tested two distinct correlations. One is an immediate correlation without a time delay, and the other is the maximum correlation with a time delay. Both were measured based on the cross-correlation of the GDP time series using \texttt{scipy.signal.correlate}~\cite{plerou1999universal}. We use the time-delayed correlation with the highest value within $\pm 10$ years of delay.

We observed positive correlations between the GDP and world trade networks (Fig.~\ref{fig:Correlation}). We observed $p$-values of less than $10^{-10}$ for all relationships; therefore, such values can be considered significant. In particular, long-range interactions reflected in the PPR exhibited a higher correlation coefficient with a raw GDP correlation (Pearson $r = 0.2343$ and Spearman $r = 0.2809$, as shown in Fig.~\ref{fig:Correlation}(a)) than the direct trade volume reflected in the normalized trade volume (Pearson $r = 0.1682$ and Spearman $r = 0.2556$, as shown in Fig.~\ref{fig:Correlation}(c)). This pattern also holds for time-delayed GDP correlations. Long-range interactions (PPR) exhibited stronger correlations than the trade volume (Pearson $r = 0.1529$ and Spearman $r = 0.2529$, as shown in Fig.~\ref{fig:Correlation}(d)). In other words, two countries with strong trade ties are more likely to experience similar fluctuations in GDP. Nonetheless, this is a natural phenomenon because all of these interconnections are in some way related to an economic exchange, which ultimately results in such fluctuations. 

Intriguingly, both immediate and time-delayed correlations with GDP are more explanatory for a PPR influence network. In other words, in terms of the GDP, adding indirect impacts is more informative than simple direct impacts in distinguishing significant countries. In addition, the magnitude of the correlation coefficient has a greater impact on the immediate GDP effect than on the value under optimal synchronization, whether indirect or direct. It can be hypothesized that, given sufficient time, countries with a weak relationship spread their economic impact poorly at a particular instance, but may spread it well in the long term after two or three steps. In the following section, we examine in depth the nature of such relationships and time delay.

\subsection{Direction and Time Delay}

\begin{figure}[t!] 
\centering 
\centerline{\includegraphics[width=\textwidth]{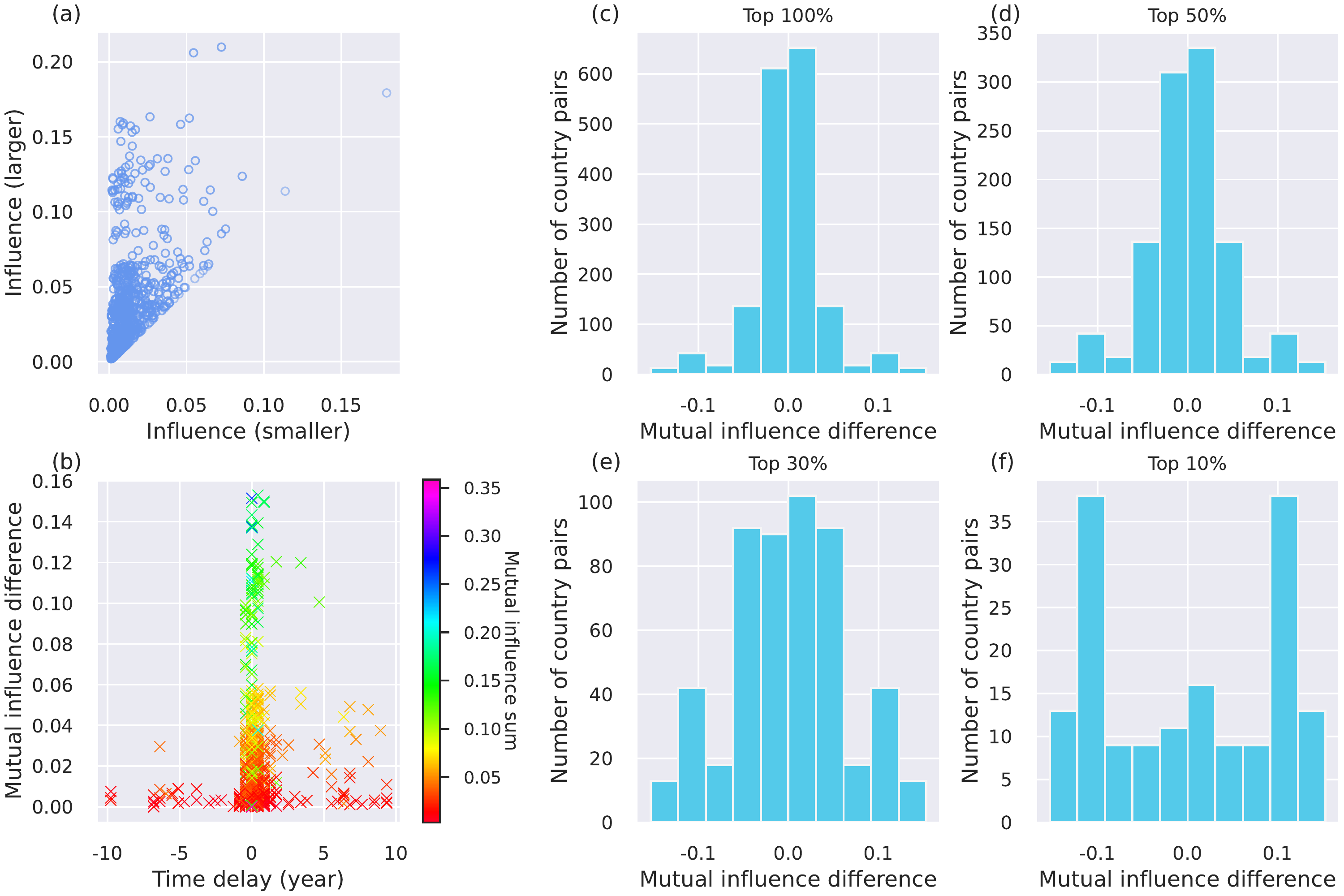}}
\caption{Disparities in mutual influence and time lag. (a) Bilateral correlation of influences as determined through the PPR. (b) Correlation between a time delay for the maximum correlation and the mutual influence. Distribution of bilateral differences in mutual influences: (c) all pairs, (d) top 50\%, (e) top 30\%, and (f) top 10\% regarding the total mutual influence of the pair of countries.}\label{fig:Direction}
\end{figure}

Understanding the directionality of systems governed by complex networks is crucial~\cite{lee2021uncovering}. For example, a weight imbalance may have a substantial effect on a dynamic system~\cite{dorfler2013synchronization, eidum2020modeling}. In this section, we examine the effects of a weight imbalance on the world trade network. We employ PPR as the primary relationship between two countries because of its higher explanatory power than the normalized trade volume, as shown in Section~\ref{subsec:correlations}.

The probability that a random walker from node $A$ (source) visits node $B$ (target) is defined as the PPR from node $A$ to $B$ and vice versa. Consequently, the PPR of both directions can differ, and the probabilities do not need to be equal. Indeed, we found that the relationships are asymmetric (Fig.~\ref{fig:Direction}(a)), indicating the existence of hidden dependencies in the weighted network~\cite{lee2021uncovering}. We may consider the effect of this asymmetry on the synchronization of the global economy. We examined the distinction between mutual influence and the optimal time delay employed in the time-delayed correlation shown in Fig~\ref{fig:Direction}(b). We observed that the time delay between the two countries is typically greater when their influence is relatively symmetrical. In addition, the greater the difference in influence, the greater the sum of the influence itself (Fig~\ref{fig:Direction}(b)--(f)). In summary, the direction of economic propagation is predominantly from larger-impact countries to smaller-impact countries, which shows a hidden dependency between countries~\cite{lee2021uncovering}.

\subsection{Mechanistic model of risk propagation}
Our empirical analysis presented in the preceding sections demonstrates that 1) the global economy is correlated, and 2) it propagates through the global trade network. In addition, we demonstrated that the total influence measured by the PPR has stronger explanatory power than the trade volume itself (see Fig.~\ref{fig:Correlation}). To understand the mechanism underlying the observed correlations, we accounted for the global economic exchange using a simple agent-based model~\cite{otto2017modeling}. We hypothesize that the current higher correlation between the PPR and GDP can be replicated by the exchange of the world trade network. Thus, we first constructed a global trade network consisting of raw trade volumes between countries. First, we assumed that the GDP of a country is more likely to be engaged if the country has more trade~\cite{otto2017modeling}. Second, the random null model predicts a proportional change in trade volume based on the size of the economy of the country. 

\begin{figure}[h!] 
\centering 
\centerline{\includegraphics[width=\textwidth]{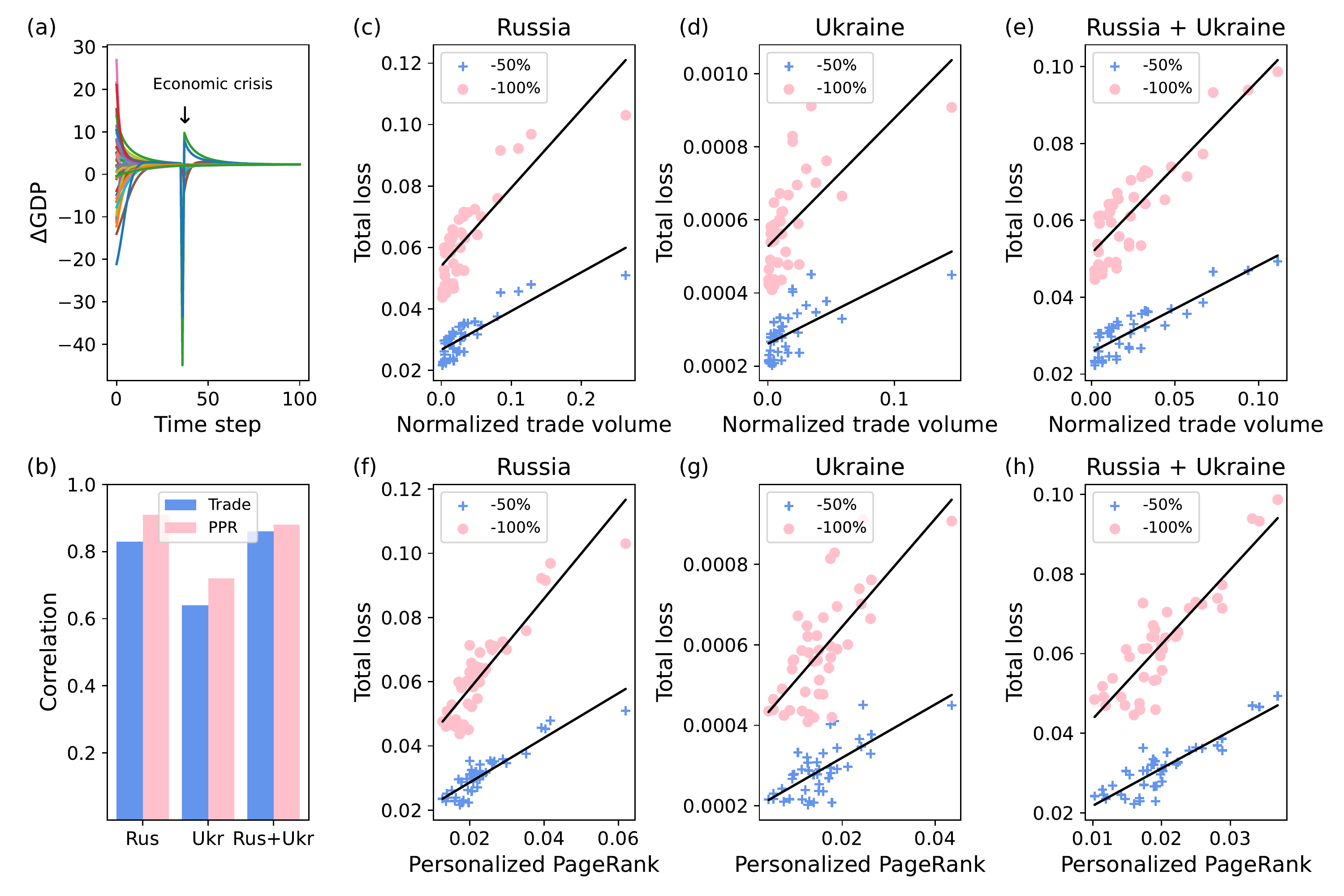}}
\caption{Agent-based modeling for the Russia-Ukraine war of 2022. (a) An illustration of how the economic crisis is implemented. We immediately reduce the GDP of Ukraine and/or the Russian Federation by a certain amount at the 35th simulation step. (b) Pearson correlations between world trade networks and the total loss in GDP. The total loss in GDP is correlated with (c)--(e) the normalized trade volume and (f)--(h) PPR. We examined three scenarios in which only the Russian GDP decreased ((c) and (f)), only the Ukrainian GDP decreased ((d) and (g)), and the GDP of both Russia and the Ukraine decreased ((e) and (h)). Here, the interdependence on the horizontal axis is the sum of the in- and out-weights from countries with an instance of GDP loss (Ukraine and/or Russian Federation) to a target country. For (e) and (h), we used either the sum of the influences for both Russia and Ukraine from/to a certain country, or the mean value of the mutual influences between the source and other countries.}
\label{fig:Simulator} 
\end{figure}

Using these factors, we constructed a mechanistic model of economic propagation. Imports, exports, and domestic consumption are the three factors considered to influence the economic growth of a country. Thus, the GDP of country $i$ is updated as follows for every simulation step:

\begin{equation}\label{equation}
P_{i,t+1} = P_{i,t} + \sum (\frac{P_{j,t}}{P_{j,0}}T_{ij} + \frac{P_{i,t}}{P_{i,0}}T_{ji}) + k P_{i,t} D_{i},
\end{equation}

\noindent where $P_{i, t}$ is the GDP of country $i$ at time $t$, and $T_{ij}$ is the raw trade volume from country $i$ to $j$. In addition, $D_i$ is the degree of domestic economic activity, estimated using the ratio of exports and imports to the GDP. We use a value of $1-R_{ex}(2010)$, where $R_{ex}(2010)$ is the ratio of exports to GDP in 2010 from KOSIS (see the Methods section). Here, $k$ is a tunable parameter that adjusts for the impact on the domestic economy. We used $k=0.015$. Owing to its absence in this dataset, we assume that the domestic economy of Ukraine is equal to that of Latvia, which is the most similar country in terms of economic scale and trade volume. At the initialization step, we assigned a country’s GDP as the average between 1990 and 2020. We consider an extreme economic event, i.e., the Russia-Ukraine war-induced economic crisis of 2022, because typical fluctuations are synchronized at a global scale, as demonstrated in the previous sections. We implemented the crisis by instantly decreasing the GDPs of the source country (or countries) by a certain amount. We examined the following three scenarios at the beginning of the economic crisis: 1) only the Russian economy, 2) only the Ukrainian economy, and 3) both economies (Fig.~\ref{fig:Simulator}(a)) are initially damaged. We then calculated the absolute value of the total loss in GDP, which is defined as the total sum of the changes in GDP after the economic crisis.

We previously demonstrated a correlation between economic influence and the decline in GDP for different countries. The outcome of our simple model was consistent with the observed data. As an indicator of how accurately the association pattern has been reproduced by our model, we observed a strong positive Pearson correlation between world trade networks (i.e., both the PPR and normalized trade volume) and the loss of GDP (Fig.~\ref{fig:Simulator}(b)--(h)). We discovered an association pattern similar to our previous empirical observations. In particular, we observed a stronger Pearson correlation of the GDP loss with the PPR ($=0.91$), followed by a correlation with the normalized trade volume ($=0.83$) for the first scenario in which the war damages only the Russian economy (Figs.~\ref{fig:Simulator}(b) and ~\ref{fig:Simulator}(c) and (f)). This also holds true for scenarios 2 and 3, with correlations of $0.72$ (PPR) to $0.64$ (normalized trade volume) for scenario 2, and $0.88$ (PPR) to $0.86$ (normalized trade volume) for scenario 3 (Figs.~\ref{fig:Simulator}(b), (d) and (e) along with (g) and (h)). Because economic synchronization occurs through multiple routes, the current state of association is the result of an accumulated exchange process through multiple routes beyond simple trade volumes. Our model is a minimalist implementation that employs a single exchange route, i.e., the global trade network, and ignores the type of goods. However, this implies that long-term trade interactions can produce the current state of GDP engagement.

\section{Discussion}
This study investigated global economic engagement through a world trade network, emphasizing the significance of indirect and long-range influences beyond direct trade volumes. The economy of a country is driven not only by the action of its domestic market but also by the cooperation of many other countries. To enhance the validity of our approach, we propose that a more comprehensive analysis of mutual influence between countries may be required. For instance, extending our analysis to a more subdivided world trade network by the types of goods, such as natural resources, agricultural products, industrial products, and services, could be a good subject for a future study. In addition, our model is rather deterministic and lacks randomness in the update rule; therefore, a refinement of the presented model is left for future research.

Many economic agents have agonized over the unpredictability and repercussions of economic crises. The recent Russian-Ukraine war has significantly impacted the global economy, increasing the cost of resources. For example, the price of oil has increased abruptly, which has affected the cost of various industries compared to before the war~\cite{adekoya2022does}. Under such a situation, determining which country has the most significant impact on the economy of a given country is crucial for economic security. Our research indicates that a simple trade volume may be insufficient to identify the most important trading partners owing to the complex nature of the economy. In addition to the PPR presented in this study, there are various ways to measure long-range impacts. Graph embedding techniques~\cite{xu2021understanding}, which have recently become popular, can be used to measure the influence between countries, and more accurate information can be estimated using a multilayer network or knowledge graph that handles multiple relationships simultaneously~\cite{johnson2022knowledge}. When an economic crisis is anticipated as a result of COVID-19, we hope that the combination of our method and precise mathematical modeling will serve as a driving force to overcome the current and other forthcoming crises.

\section{Author contributions}
All authors designed the study and wrote the manuscript. Sungyong Kim collected and analyzed the data.

\begin{acknowledgments}
We thank Dr. Taekho You for the valuable and constructive suggestions. This research was supported by the Korea Institute of Science and Technology Information (KISTI). KISTI also supported this research by providing access to KREONET, a high-speed Internet connection. The funders played no role in the study design, data collection and analysis, decision to publish, or preparation of the manuscript. 
\end{acknowledgments}

\bibliography{reference.bib}

\end{document}